\documentclass[a4paper,fleqn,usenatbib,letter]{mnras}

\usepackage[T1]{fontenc}
\usepackage{ae,aecompl}

\usepackage{graphicx}	
\usepackage{amsmath}	
\usepackage{amssymb}	

\title{Testing the cores of first ascent red-giant stars using the period spacing of g modes}

\author[Lagarde N. et al.]{
Lagarde N.,$^{1}$\thanks{E-mail: lagarde@bison.ph.bham.ac.uk}
Bossini D.,$^{1}$, Miglio A.$^{1}$, Vrard M.$^{2}$ and Mosser B.$^{2}$\\
$^{1}$School of Physics and Astronomy, University of Birmingham, Edgbaston, Birmingham B15 2TT, UK\\
$^{2}$ LESIA, Observatoire de Paris, PSL Research University, CNRS, Universit\'e Pierre et Marie Curie, Universit\'e Paris Diderot, \\92195 Meudon, France
}

\date{Accepted 2015 December 10. Received 2015 December 10; in original form 2015 November 11}

\pubyear{2015}

\begin{document}
\label{firstpage}
\pagerange{\pageref{firstpage}--\pageref{lastpage}}
\maketitle

\begin{abstract}  
In the context of the determination of stellar properties using asteroseismology, we study the influence of rotation and convective-core overshooting on the properties of red-giant stars.
We used models in order to investigate the effects of these mechanisms on the asymptotic period spacing of gravity modes ($\Delta \Pi_1$) of red-giant stars that ignite He burning in degenerate conditions (M$\lesssim$2.0 M$_{\odot}$). We also compare the predictions of these models with \textit{Kepler} observations.
For a given $\Delta\nu$, $\Delta \Pi_1$ depends not only on the stellar mass, but also on mixing processes that can affect the structure of the core. We find that in the case of more evolved red-giant-branch (RGB) stars and regardless of the transport processes occurring in their interiors, the observed $\Delta \Pi_1$ can provide information as to their stellar luminosity, within $\sim$10-20\%.
In general, the trends of $\Delta \Pi_1$ with respect to mass and metallicity that are observed in \textit{\textit{Kepler}} red-giant stars are well reproduced by the models. 

\end{abstract}




\section{Introduction}

In recent years, a great amount of asteroseismic data has been obtained by the CoRoT and \textit{\textit{Kepler}} space missions for a large sample of red giants \citep[e.g.,][]{deRidder09,Borucki10}. One of the most important breakthroughs resulting from these missions is the detection of non-radial mixed-mode oscillations in red giants, which contribute to the determination of the precise evolutionary phases of giant stars \citep[e.g.,][]{Bedding11,Stello13,Mosser14}. 
This adds invaluable and independent constraints to stellar models. In particular, the comparison between models that include a detailed description of transport processes in stellar interiors and asteroseismology, in addition to spectroscopy, opens a promising new path for developing our understanding of stars \citep[e.g.][]{Lagarde15a}.
Indeed, asteroseismology has also been widely used to estimate stellar properties (e.g. stellar mass, radius, and distance), providing a fundamental contribution to the characterization of planet hosting stars \citep[e.g.][]{Huber13a, Johnson14} and stellar populations \citep{Miglio13}.    
Rotation and convective-core overshoot are two examples of a number of key processes that change all the outputs of stellar models, with a significant impact on asteroseismic observables \citep[e.g.][]{Eggenberger10,Lagarde12a,Montalban13,Bossini15,Constantino15}. Consequently, transport processes can have a significant impact on the determination of planet and star masses.
In this paper, we focus on the effects of rotation and overshooting on the asteroseismic modelling of first-ascent red giants (M$\leq$2.0 M$_{\odot}$) that ignite He burning in degenerate conditions by comparing stellar models that include shellular rotation or overshoot to non-rotating models. We examine the asymptotic period spacing of gravity modes, $\Delta\Pi_1$, probing the stellar interiors. The influence of rotation and overshoot on the determination of the global properties of red-giant stars are investigated along the red giant branch using two evolutionary codes. A comparison with the observations of the \textit{Kepler} mission is presented. 

\section{Stellar evolution models}
\label{models}
\subsection{Physical inputs }
		
Stellar models are computed using STAREVOL \citep[e.g.][]{Lagarde12a} and adopting solar metallicity (with the chemical composition by \citealt{Asplund09}) for a range of masses between 1.0 and 2.0 M$_{\odot}$. In order to quantify the impact of each transport process at the various evolutionary phases, we computed models by adopting the following assumptions: (1) standard models (no mixing mechanism other than convection); (2) models that include overshoot with a parameter d$_{\rm{over}}$/H$_{p}$ of 0.2 at the Schwarzschild's border \citep{Maeder75}; (3) models including rotation-induced mixing with $V_{\rm{ZAMS}}$=50 km.s$^{-1}$ (which corresponds to 13\% and 12\% of the critical velocity for 1.0 and 2.0 M$_{\odot}$, respectively) and $V_{\rm{ZAMS}}$=250 km.s$^{-1}$ ($\sim$0.62$V_{\rm{crit}}$ for 1.8-2.0 M$_{\odot}$ stars). These models include the transport of angular momentum and chemicals by meridional circulation and shear turbulence, following the formalism of \citet{Zahn92} and \citet{MaZa98}. A more detailed description of the model used to describe rotation can be found in \citet{Lagarde12a}. We also computed two sets of stellar models with alpha-enrichment ([$\alpha$/Fe]=0.21 and 0.15; Sect.4). 

\subsection{From the H-burning phase to the RGB}

\begin{figure} 
	\centering
		\includegraphics[angle=0,width=0.43\textwidth, clip=true,trim=1.3cm 2.5cm 2cm 2cm]{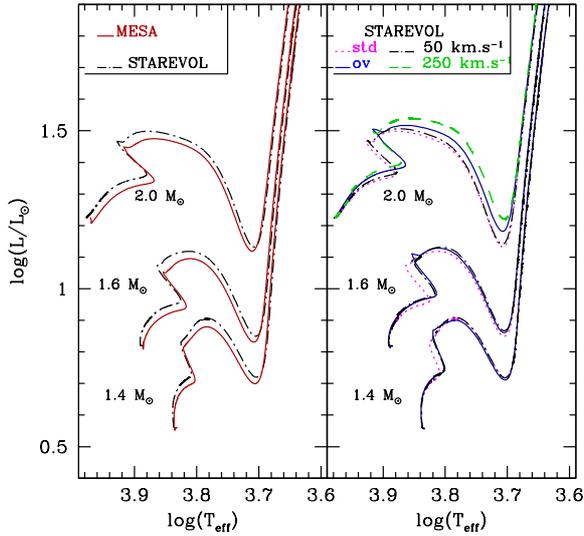}
			  \caption{Theoretical evolution tracks in the Hertzsprung-Russel (HR) diagram for models at 3 stellar mass (1.4, 1.6, and 2.0 M$_{\odot}$) at solar metallicity. In the left panel, the tracks are shown for standard models computed using the STAREVOL and MESA codes (dash-dotted and solid lines, respectively) from the zero-age main sequence. The right panel shows models computed using STAREVOL that include rotation-induced mixing effects (at two velocities), overshooting, and standard models.}
	\label{Compa_HR}
\end{figure}
In the set of models that were constructed, it is possible to distinguish between two evolutionary regimes on the sub-giant branch, depending on the core structure and mass \citep{KiWe90}: 
(1) stars with $M$$\lesssim$1.8 M$_{\odot}$ (hereafter referred to as low-mass stars; LMS) have a core mass fraction smaller than the Sch\"onberg-Chandrasekhar limit (SCL) at the end of the central H-burning phase. Then H burning continues in a thick shell until the core mass reaches the SCL. This phase may last for a significant fraction of the H-burning phase. 
(2) Stars with masses $M$$\gtrsim$1.8 M$_{\odot}$ (hereafter referred to as intermediate-mass stars; IMS) have a core mass greater than the SCL at the end of the main sequence, and the core contracts at the Kelvin-Helmoltz timescale. This limit in the mass can change with metallicity and transport processes.
These stars then immediately experience central contraction, which is accompanied by envelope inflation and brings the star to the red giant stage. 
This distinction was also apparent in the $\Delta\Pi_1$-$\Delta \nu$ diagram of the red giants observed by the \textit{\textit{Kepler}} mission \citep{Mosser14,Stello13}.

\subsection{The asymptotic period spacing of g modes}
\begin{figure} 
	\centering
		\includegraphics[angle=0,width=0.46\textwidth, clip=true,trim=0.8cm 0.7cm 1.2cm 1cm]{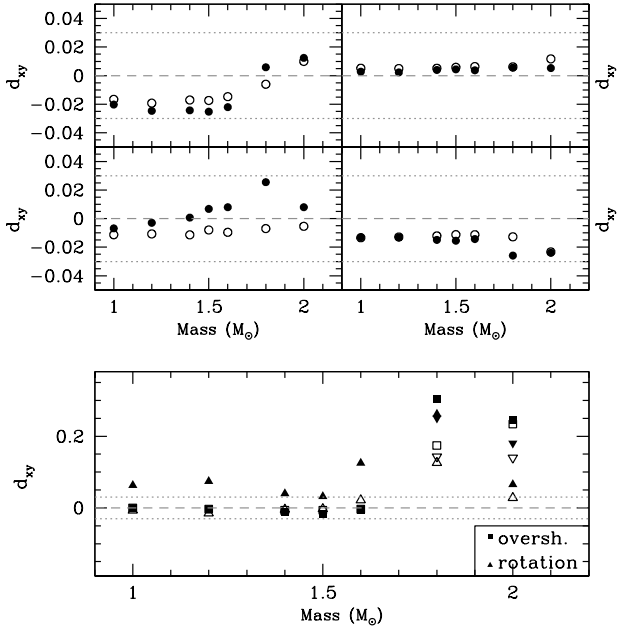}
			  \caption{Effects of various sources of uncertainty on $\Delta\Pi_{1}$ ($d_{xy}=(\Delta \Pi_{y}-\Delta \Pi_{x})/\Delta\Pi_{x}$) as a function of stellar mass. The top panels show the effects of the stellar evolution codes, STAREVOL ($\Delta\Pi_{S}$) and MESA ($\Delta\Pi_{M}$, top-left); a variation of +0.15 $\mu$Hz in the large separation (top-right); metallicity [Fe/H]=0 and -0.2 (bottom-left); and a variation +100 K in the effective temperature (bottom-right). Relative differences are shown at two large separations ($\Delta \nu$=15 and 6 $\mu$Hz, with filled and empty symbols, respectively) for standard models. The bottom panel shows the effects of rotation (triangles) and overshooting (squares) on $\Delta\Pi_{1}$. Two initial velocities are shown: $V_{\rm{ZAMS}}$ = 50 (upwards triangles) and 250 km.s$^{-1}$ (downwards triangles).}
	\label{Compa_deltaPi}
\end{figure}

The period spacing of dipolar g modes can be approximated by the asymptotic relation \citep{Tassoul80}: $\Delta\Pi_1 =\sqrt{2} \pi^{2}/\int_{r_{1}}^{r_{2}} N \frac{dr}{r},$where $N$ is the Brunt-V\"ais\"al\"a frequency and $r_{1}$ and $r_{2}$ define the domain (in terms of the radius) where g modes are trapped. 
Therefore, $\Delta\Pi_1$ provides information about the detailed properties of the stellar core and has a unique capability to constrain stellar models \citep[][]{Mosser12a, Montalban13,Lagarde12a}.
Moreover, the observed values of $\Delta\Pi_1$ and $\Delta \nu$ are frequently used to obtain a first estimate of stellar mass and age from stellar models \citep[e.g.][]{Deheuvels12,Johnson14,Martig15}.
In order to test the reliability of these determinations, we investigate various sources of uncertainties (see Fig.\,\ref{Compa_deltaPi}) that can affect the predicted $\Delta\Pi_1$ at a given $\Delta \nu$: 

\begin {itemize}
\item We use the MESA stellar evolution code \citep[e.g.,][]{Paxton11} to compute standard and overshooting models at solar metallicity, within the same mass range and with the same chemical composition \citep{Asplund09} as was used in STAREVOL. 
Despite the fact that the two codes adopt different prescriptions (e.g. the equation of state and nuclear reaction rates), we note that when similar mixing prescriptions are adopted, the differences in $\Delta\Pi_1$ are less than 2.5\% for LMS and less than 1\% for IMS (Figs.\ref{Compa_HR}\&\ref{Compa_deltaPi}).

\item An uncertainty of $\pm$0.15 $\mu \rm{Hz}$ on $\Delta \nu$ implies approximately 1\% difference in $\Delta\Pi_1$ (top-right panel of Fig.\,\ref{Compa_deltaPi}). 

\item A change of $\pm$0.2 dex (the typical uncertainty on the metallicity) on [Fe/H] (see middle-left panel of Fig.\,\ref{Compa_deltaPi}) has a minor impact on the theoretical $\Delta\Pi_1$ (less than $\sim$2\% for the whole mass range). 

\item We then investigate the effects of a typical error of 100 K on the $T$$_{\rm{eff}}$ due to either observational uncertainties or theoretical predictions (for further details, see \citealt{Cassisi14} and references), obtaining a change in $\Delta\Pi_1$ of less than 2\% for LMS and 3\% for IMS (Fig.\,\ref{Compa_deltaPi}, middle-right panel). 

\end{itemize}

All these sources of uncertainty are smaller than the effects of rotation and overshooting (see the bottom panel of Fig.\,\ref{Compa_deltaPi}), especially if we consider IMS ($\sim$10\% on $\Delta\Pi_{1}$). In addition, when the velocity at the zero-age main sequence (ZAMS) increases, the relative difference in $\Delta\Pi_{1}$ increases ($\sim$6\% at $V_{\rm{ZAMS}}$=50 km.s$^{-1}$ and $\sim$18\% at $V_{\rm{ZAMS}}$=250 km.s$^{-1}$ for a 2.0 M$_{\odot}$ star at $\Delta\nu$=15 $\mu$Hz).
For the following discussion, we focus on solar metallicity.

\section{Determining the properties of RGB stars from $\Delta \Pi_{1}$ }
\label{deltaPi}

\subsection{Standard models}

\begin{figure} 
	\centering
		\includegraphics[angle=0,width=0.43\textwidth, clip=true,trim=0cm 0.6cm 1cm 1cm]{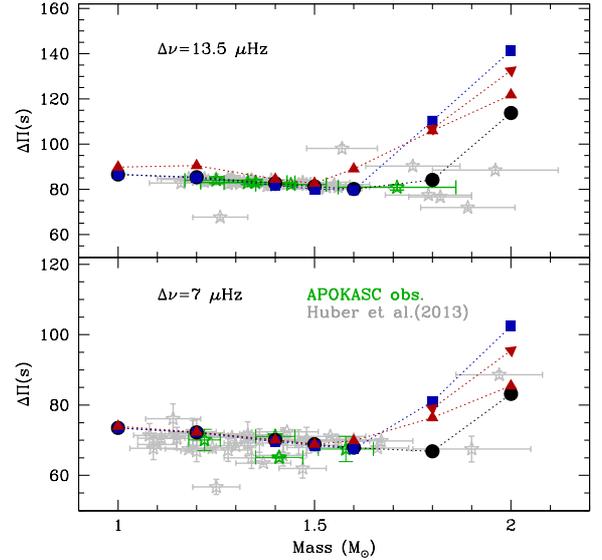}
			  \caption{$\Delta\Pi_{1}$ at $\Delta \nu$=13.5$\mu$Hz (top panel) and at $\Delta \nu$=7$ \mu$Hz (bottom panel) as a function of the initial stellar mass, for models that follow standard predictions (black circles), include overshooting on the main sequence (blue squares), or include the effects of rotation-induced mixing, with $V_{\rm{ZAMS}}$=50 km.s$^{-1}$ (red upwards triangles) and $V_{\rm{ZAMS}}$=250 km.s$^{-1}$ (red downwards triangles). \textit{\textit{Kepler}} observations of red giants from Vrard et al. (2015, submitted to  A\&A) are indicated using grey and green stars. We distinguish \textit{Kepler}-observed red giants with metallicity determinations obtained by APOGEE \citep[][green stars]{Pinsonneault14} and \citet[][grey stars]{Huber14}.}
	\label{deltaPiq}
\end{figure}

\begin{figure} 
	\centering
		\includegraphics[angle=0,width=0.46\textwidth, clip=true,trim=0.2cm 3.7cm 1cm 1.5cm]{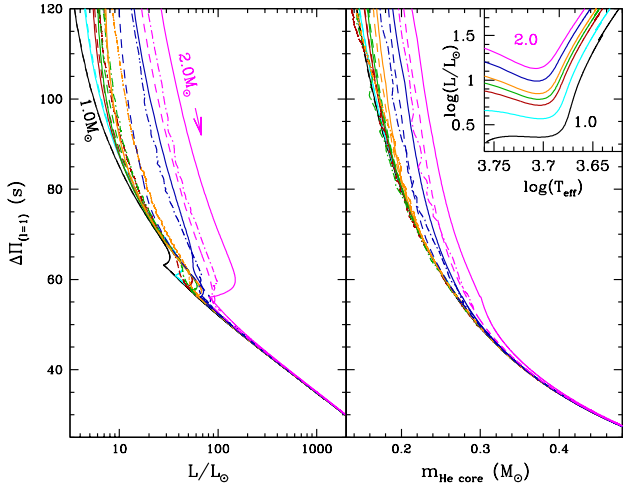}
			  \caption{The evolution of $\Delta\Pi_{1}$ during the RGB with the stellar luminosity (left panel) and mass of the helium core (right panel). Models are computed at solar metallicity in the mass range between 1.0 and 2.0\,M$_{\odot}$: (from left to right) 1.0\,M$_{\odot}$ (black), 1.2\,M$_{\odot}$ (cyan), 1.4\,M$_{\odot}$ (red), 1.5\,M$_{\odot}$ (green), 1.6\,M$_{\odot}$ (yellow), 1.8\,M$_{\odot}$ (blue), and 2.0\,M$_{\odot}$ (magenta), following standard prescriptions (dashed lines), including overshooting (solid lines), and including rotation (dash-dotted lines). The panel inserted on the right of the figure shows the HR diagram.}
	\label{Hecore}
\end{figure}

Figure \ref{deltaPiq} (upper panel) shows $\Delta\Pi_1$  as a function of stellar mass, for standard prescriptions (filled black circles) at $\Delta \nu$=13.5 $\mu$Hz. 
$\Delta \Pi_1$ shows almost the same value for stars with masses lower than $\sim$1.6 M$_{\odot}$, while it begins to differ at higher masses. The higher the mass, the later (in terms of luminosity) the stars reach similar central conditions in terms of density and temperature. Consequently, the asymptotic period spacing of g modes of IMS does not follow the same relation with the luminosity as that of LMS (see Fig.\,\ref{Hecore}). Figure \ref{deltaPiq} (lower panel) also shows the prescription at lower $\Delta \nu$=7$ \mu$Hz, where the limit in the mass is located at M$\sim$1.8 M$_{\odot}$.

During the first dredge-up, the convective envelope deepens inside the star, while the degenerate He core contracts. This results in changes in the g mode cavity (in terms of its size and density), and reduces the value of $\Delta\Pi_1$. The monotonic increase of the stellar luminosity along the RGB then momentarily stops when the hydrogen-burning shell (HBS) crosses the molecular weight barrier left behind by the first dredge-up. At that moment, the mean molecular weight of the HBS decreases, which implies a decrease of the total stellar luminosity. This is referred to as the bump in the luminosity function (see Fig.\,\ref{Hecore} for the drop in the luminosity at $\Delta \Pi_1\sim$60 s). When the region of nuclear energy production has completely passed this discontinuity, the stellar luminosity increases again. During these changes in the stellar luminosity, the He-core continues to grow and then $\Delta \Pi_{1}$ decreases. Two regimes in the $\Delta \Pi_{1}$-L diagram can also be distinguished before and after the bump luminosity (Fig.\,\ref{Hecore}): 
\begin{itemize}
\item From the bump luminosity to the RGB tip, the value of $\Delta\Pi_{1}$ is directly related to the degenerate He core properties (right panel of Fig.\,\ref{Hecore}). In addition, the properties of the shell and therefore the stellar luminosity are mainly determined by the mass and the radius of the degenerate He core \citep{KiWe90}. This implies a univocal relation between $\Delta\Pi_{1}$ and $L$ (see Fig.\,\ref{Hecore}). 
\item For stars with $M$$\lesssim$ 1.7 M$_{\odot}$, this relation can also be used slightly before the bump luminosity (in the range between $\sim$12 L$_{\odot}$ and $\sim$60 L$_{\odot}$) in order to deduce the stellar luminosity from the observed value of $\Delta \Pi_1$, since their cores already satisfy the strong degeneracy condition. 
\end{itemize}

\begin{figure} 
	\centering
		\includegraphics[angle=0,width=0.38\textwidth, clip=true,trim=0.1cm 0.6cm 1cm 0.5cm]{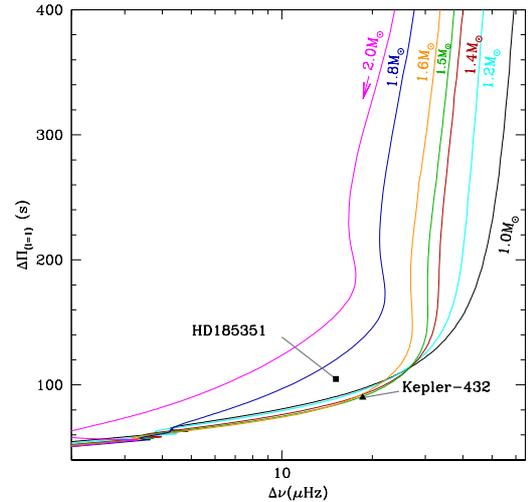}
			  \caption{Evolutionary model tracks showing $\Delta\Pi_{1}$ as a function of $\Delta \nu$. Models are computed at solar metallicity in the mass range between 1.0 and 2.0 M$_{\odot}$ (from right to left) including overshooting effects (solid line). Two \textit{\textit{Kepler}} giants are also indicated.}
	\label{Dpgrot}
\end{figure}

Since this method uses only $\Delta\Pi_{1}$, the inferred luminosity (and distance) is independent of the standard method that employs the seismic radius \citep{Miglio13}. A detailed comparison of these two methods will be useful in the case of stellar clusters or field stars that were observed by Hipparcos. This will not be discussed in this letter. 
However, we should keep in mind that at low $\Delta \nu$ (i.e. a high luminosity on the RGB), when gravity-dominated mixed modes have high inertias, detecting $\Delta\Pi_{1}$ becomes challenging \citep{Grosjean14}. 
Prior to this stage, the effects of different transport processes (see Sect.3.2 and Figs.\ref{Hecore}\&\ref{Dpgrot}) should be taken into account while determining stellar mass using the observed $\Delta\Pi_1$ and $\Delta\nu$ \citep[e.g.][]{Deheuvels12, Johnson14}.

\subsection{Rotating and overshooting models}

The effects of rotation on the size of the He core at the end of the main sequence are similar to the effects of overshooting. During the main sequence, rotation-induced mixing brings fresh hydrogen fuel into the longer-lasting convective core and transports the H-burning products outwards. This results in a more massive helium core at the turnoff than in the standard case. This effect also shifts the tracks toward higher effective temperatures and luminosities \citep[e.g.][]{Ekstrom12}, as well as larger $\Delta\Pi_{1}$ at given $\Delta\nu$ throughout their evolution (Fig.\,\ref{deltaPiq}). 
Rotating and overshooting models behave like stars with higher masses throughout their evolution.
However, overshooting models cannot mimic the impact of rotation on stellar nucleosynthesis. In fact, rotation modifies the internal and surface chemical abundances during the main sequence \citep[e.g.][]{Palacios06, ChaLag10}, which results in smoother chemical profiles. As a consequence, at a given $\Delta \nu$, the behaviour of the Brunt-V\"ais\"al\"a frequency differs from those of the standard and overshooting models. Because of changes in the mean molecular weight, the Brunt-V\"ais\"al\"a frequency is slightly larger in rotating models. 
Since a 1.0 M$_{\odot}$ star has no convective core during the main sequence, overshooting plays no role, contrary to rotation that changes its chemical properties. This explains why models with rotation exhibit an impact on $\Delta\Pi_1$, as shown in Fig.\,\ref{deltaPiq}.
For the same reasons, at a given $\Delta \nu$ in Fig.\,\ref{deltaPiq}, the stellar mass for which $\Delta\Pi_1$ is at its lowest value changes with rotation. 
At the end of the first dredge-up, the convective envelope removes the changes in the Brunt-Va\"is\"al\"a frequency that were induced by rotation. 
The measurement of $\Delta\Pi_1$ in red-giant stars brighter than the bump luminosity can give an accurate measurement of their stellar luminosity, regardless of the physical processes occurring on the stellar interior at this stage (see Fig.\,\ref{Hecore}). This also provides the He core properties (mass and radius). 
Assuming an uncertainty of $\pm$2 s on $\Delta\Pi_1$, the typical uncertainty of the luminosity determination increases along the RGB, from 9\% for LMS, before the bump luminosity, to $\sim$25\% immediately after the bump luminosity. 

\section{Comparison with observations}

\begin{figure} 
	\centering
		\includegraphics[angle=0,width=0.4\textwidth, clip=true,trim=1.6cm 0.6cm 1.5cm 1cm]{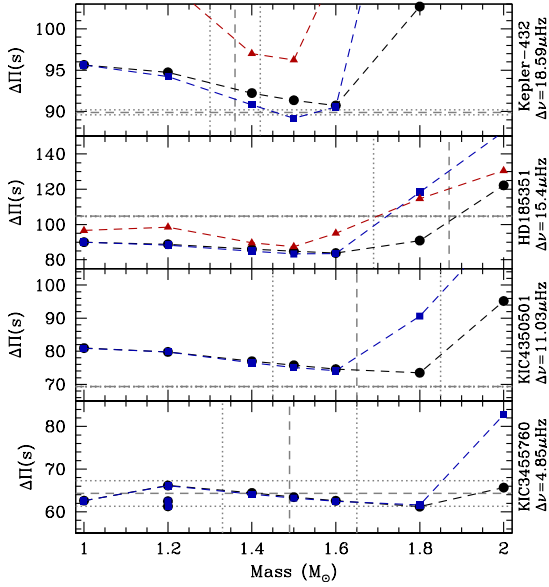}
			  \caption{Comparison between predicted and observed $\Delta\Pi_1$ for four cases: \textit{Kepler}-432, HD185351, KIC4350501, and KIC4355760. The horizontal dashed line shows the observed value of $\Delta\Pi_{1}$ and its error (dotted lines). The vertical dashed line shows the value of the stellar mass that was deduced from the scaling relations. The theoretical $\Delta\Pi_{1}$ following standard prescriptions (black filled circles), including the effects of overshooting (blue filled squares), and rotation (red filled triangles) are also shown. } 
	\label{Dpgeachstar}
\end{figure}

Figure \ref{deltaPiq} shows a comparison of the model predictions with the \textit{\textit{Kepler}} observations of Vrard et al. (submitted). We select observations within a 0.2 dex bin in [Fe/H] arround the solar metallicity, and 0.5 $\mu$Hz in $\Delta\nu$. Although IMS RGB stars are not numerous, our models reproduce well the observations for the whole mass range. The lack of observations at higher masses prevents any confirmation of the dispersion of $\Delta\Pi_1$ for IMS that was predicted by the rotating and overshooting models. For the case of LMS, models reproduce the observed $\Delta\Pi_1$, independent of the transport process.
In Fig.\,\ref{Dpgeachstar}, we compare our theoretical predictions with four stars that were observed by the \textit{Kepler} mission. The observed $\Delta\Pi_1$ and the stellar mass obtained from scaling relations from $\Delta\nu$ and the frequency of maximum oscillation power, $\nu_{max}$ ($M$$_{\rm{SR}}$) are available.
\begin{itemize}
\item \textit{Kepler}-432 is an evolved star ($\Delta\Pi_1=89.9\pm$0.3 s; $\Delta\nu$=18.59 $\mu$Hz; $M_{\rm{SR}}$=1.36$\pm$0.06 M$_{\odot}$) and hosts three planets \citep{Quinn14}. In our computations, standard and rotating models could not reproduce these observations. On the other hand, overshooting models appear to show a possible agreement if we consider a stellar mass that is between 1.45 M$_{\odot}$ and $\sim$1.55 M$_{\odot}$. The lower limit ($\sim$1.45 M$_{\odot}$) is consistent with the mass that was deduced from the scaling relations.

\item HD185351 is a giant star ($\Delta\Pi_1$=104.7$\pm$0.2 s; $\Delta\nu$=15.4 $\mu$Hz; $M_{\rm{SR}}=1.87\pm$0.18 M$_{\odot}$) and hosts giant planets \citep{Johnson14}. The standard model provides a mass very close to $M$$_{\rm{SR}}$, while the rotating and overshooting models provide a lower value ($M\approx$1.7 M$_{\odot}$). 
These determinations are consistent with a mass that is higher than 1.5 M$_{\odot}$, as was underlined by \citet{Johnson14}. However, we need more information about the surface and the core-rotation rates in order to accurately compare observations with rotating models. This case shows the importance of rotation (or overshooting) in the determination of stellar mass with $\Delta\Pi_1$ and $\Delta \nu$, for early-RGB stars.

\item KIC4350501 ($\Delta\Pi_1=69.3\pm$0.1 s; $\Delta\nu$=11.03 $\mu$Hz; $M$$_{\rm{SR}}=1.65\pm0.2$ M$_{\odot}$) and KIC3455760 ($\Delta\Pi_1=64.3\pm$3 s; $\Delta\nu$=4.85 $\mu$Hz; $M$$_{\rm{SR}}=1.49\pm$0.16 M$_{\odot}$) are APOKASC red-giant stars that were studied by \citet{Martig15} in the context of young $\alpha$-rich stars. We computed specific models representative of the metallicity of these two stars at [Fe/H]=$-$0.10$\pm$0.03 with [$\alpha/$Fe$]=0.21\pm$0.05 for KIC4350501; and at  [Fe/H]=0.01$\pm$0.03 with [$\alpha/$Fe$]=0.15\pm$0.04 for  KIC3455760. In the case of KIC4350501, we note that our standard and overshooting models cannot reproduce the low observed values of  $\Delta\Pi_1$. Because of its very small uncertainty, we cannot obtain mass limits from our comparisons. The $\Delta\Pi_1$ and $\Delta\nu$ of KIC3455760 cannot provide a unique mass determination. 
\end{itemize}

\section{Conclusions}

We investigated the effects of rotation-induced mixing and overshooting on the theoretical determination of $\Delta\Pi_{1}$ for LMS and IMS ($M\leq$2.0 M$_{\odot}$). Contrary to the effect of overshooting, rotation-induced mixing has an impact on the value of $\Delta\Pi_{1}$ for stars without a convective core during the main sequence (for 1.0 M$_{\odot}$ at $\Delta \nu$=15 $\mu$Hz, $\Delta\Pi_{1}$ is $\sim$6 s longer). 
We evaluated the important effects of rotation and overshooting in terms of the determination of stellar mass using $\Delta\Pi_{1}$ and $\Delta\nu$. Determinations can differ up to a value of 0.2 M$_{\odot}$. We ascertained that in certain cases, $\Delta\Pi_1$ and $\Delta\nu$ do not provide a unique solution for the stellar mass. 
Regardless of the transport processes occurring in their interior, the observed $\Delta\Pi_{1}$ for RGB stars above the bump luminosity provides an indirect measurement of the stellar luminosity (and therefore, the distance), with an uncertainty between $\sim$9\% and 25\% before and after the bump luminosity, respectively, as well as information on the properties of their degenerate He cores. Our predictions are confirmed by comparison with very recent measurements of $\Delta\Pi_{1}$ for \textit{Kepler}-observed red-giant stars (Vrard et al., submitted). 

As accurate determinations of stellar properties are needed in other domains of astrophysics, such as those concerning exoplanets or galactic studies, it is crucial to take into account the effects of transport processes in order to improve the accuracy of the estimation of stellar properties from seismic observations.

\section*{Acknowledgements}
NL acknowledges financial support from the Marie Curie Intra-European fellowship (FP7-PEOPLE-2012-IEF). 



\bibliographystyle{mnras}
\bibliography{Reference}

\bsp	
\label{lastpage}
\end{document}